\documentstyle[preprint,aps,prc]{revtex}
\def\pmb#1{\setbox0=\hbox{#1}%
  \kern-.02em\copy0\kern-\wd0
  \kern.04em\copy0\kern-\wd0
  \kern-.02em\raise.04em\box0 }
\def\bolddot{{\pmb{$\cdot$}}}
\def\boldpi{{\pmb{$\pi$}}}
\begin{document}
%
%
\preprint{IU/NTC 96-13, TRI-PP-97-2}
\title{The Deconfinement Phase Transition in Asymmetric Matter}
\author{Horst M\"uller\footnote{Present address:
        {\em TRIUMF, 4004 Wesbrook Mall, Vancouver, B.C. Canada V6T 2A3}}} 
\address{Physics Department and Nuclear Theory Center\\
         Indiana University,\ \ Bloomington,~Indiana\ \ 47405}
\vskip1in
\date{\today}
\maketitle

\begin{abstract}
We study the phase transition of asymmetric hadronic matter to a quark-gluon
plasma within the framework of a simple two-phase model.
The analysis is performed in a system with two conserved charges (baryon
number and isospin) using the stability conditions on the free energy, the
conservation laws and Gibbs' criteria for phase equilibrium.
The EOS is obtained in a separate description for the hadronic phase 
and for the quark-gluon plasma.
For the hadrons, a relativistic mean-field model
calibrated to the properties of nuclear matter is used, and a bag-model type
EOS is used for the quarks and gluons.
The model is applied to the deconfinement phase transition that may occur in
matter created in ultra-relativistic collisions of heavy ions.
Based on the two-dimensional coexistence surface (binodal),
various phase separation scenarios and the Maxwell construction through the
mixed phase are discussed.
In the framework of the two-phase model the phase transition 
in asymmetric matter is continuous (second-order by Ehrenfest's definition)
in contrast to the discontinuous (first-order) transition of symmetric systems.
\end{abstract}

\vspace{20pt}
%
%
\section{Introduction}
The determination of the phase structure of strongly interacting matter
is one of the major challenges in theory and experiment.
In our contribution we study the phase transition from hadrons to a 
quark-gluon plasma in asymmetric matter.
Based on a two-phase model we describe new qualitative features that arise as
a function of the isospin and that may be relevant for the phenomenological
description of highly energetic heavy-ion collisions.

The creation of the quark-gluon plasma and the elucidation of its
properties is the fundamental goal in relativistic heavy--ion physics.
Although experiments at the Brookhaven AGS and at the CERN SPS have provided
evidence that in highly energetic collisions an intermediate
state is formed which is characterized by very high energy densities
\cite{STACHEL92}, it is
not clear if this new phase of matter has actually been reached
(for a recent review, see \cite{HARRIS96}).

On the theoretical side the determination of the equation of state (EOS)
of strongly interacting matter at high temperatures and densities is a
major problem in QCD.
One of the most important issues here is revealing the phase
diagram of matter and the nature of the transition.
Most reliable results come from lattice calculations for systems at zero
baryon density (for an updated status report see \cite{KANAYA96}).  
These calculations clearly indicate a first-order transition for pure
$SU(3)$ gauge theories \cite{BOYD95}. When dynamical quarks are added the
situation is less transparent. The nature of the transition
not only depends on the number of 
flavors but also on their masses \cite{BROWN90,IKKY95}, particularly on the
mass of the strange quark. 
For example, three light flavors show a first-order transition that is
absent for larger masses \cite{BROWN90}.
Despite the need of corresponding information
in applications to heavy--ion collisions, lattice calculations at finite
baryon density presently remain at an early stage without a definite
conclusion \cite{KANAYA96}.

To the best of our knowledge, there are no lattice
calculations available for asymmetric systems, {\em i.e.}, systems
at finite baryon density and with nonzero isospin.
However, it has been demonstrated in case of the liquid-gas phase transition
of nuclear matter that asymmetric systems have qualitatively different
properties than symmetric systems \cite{BB80,G92,HMUE95}. In particular, the
transition in an asymmetric system is {\sl continuous} and of second order 
according to Ehrenfest's definition, in contrast to the discontinuous,
first-order transition in symmetric systems.
The goal of the present work is to extend these studies to the
deconfinement phase transition of asymmetric matter.
Since lattice calculations cannot provide the required information at present,
we resort to a separate description of the hadronic and the quark-gluon phase.
This approach is not new and was previously employed to investigate
the existence of quark cores in neutron stars \cite{COLLINS75,BAYM76}.
Although such a formulation oversimplifies the full complexity of the problem,
it is useful for providing a first orientation to the new features that
arise when a system with two conserved charges, {\em i.e.},
baryon number and isospin, undergoes a transition.

The EOS for the quark-gluon phase is based on a simple ``bag model'', where
the thermodynamic potential is modeled as a sum of a perturbative contribution
and a nonperturbative part which is represented by the bag constant.
Along these lines we adopt a model which is based on two massless flavors
and which includes the first-order correction in the strong coupling constant
$\alpha_s$ \cite{BAYM76,KUTI80}.

The hadronic EOS is generated in a relativistic mean-field approach
involving the interaction of Dirac nucleons with isoscalar scalar and vector
meson fields and with isovector $\rho$ meson and pion fields \cite{SW86,BDS92}.
The central object in this approach is an energy functional of the meson 
mean fields, which, in principle, can be formulated without reference
to an explicit Lagrangian \cite{FST96}.
It has been shown in numerous applications that this type of model
provides an excellent description of nuclear matter and of bulk nuclear
properties throughout the periodic table
\cite{BDS92,FST96,BOGUTA77,REINHARD86,FPW87,FURNSTAHL89,BODMER89,GAMBHIR90,%
FURNSTAHL93}.
Thus we can calibrate our hadronic equation of state at zero temperature and
normal nuclear densities, and then extrapolate into the 
regime of high density and temperature appropriate for the phase transition.
The model we study involves cubic and quartic isoscalar self-couplings.
Although nonlinear $\rho$ meson interactions are possible \cite{HMUE96},
the nuclear symmetry energy is achieved using the simplest
possible coupling of the $\rho$ meson field to the nucleon.
The pions are incorporated by adding their one-body contribution, {\em i.e.},
the contribution of an ideal Bose gas with an effective chemical potential,
to the thermodynamic potential. The effective chemical potential is chosen
such that the $\rho$ meson field couples to the total isospin density,
which receives a contribution from nucleons and pions.

At low temperatures, the qualitative features studied here depend strongly 
on the nuclear symmetry energy.
At higher temperatures, the inclusion of the pions is crucial due to
the onset of a pion condensate in very asymmetric systems.
The pions are incorporated in a very simple manner. Generally
the isovector dependence of the hadronic EOS may be more complicated than
in our model;
however, these properties are not well known, particularly when one is far
from symmetric matter.
In this context it is important to realize that the hadronic model is
formulated in terms of effective degrees of freedom.
The two conserved charges, baryon number and total isospin, are the relevant
quantities in the thermodynamic analysis whereas
the particle numbers of the individual species may change when parameters,
{\em e.g.}, temperature and pressure, vary.
In the present case the pion condensed phase resembles closely matter 
containing $\Delta$-resonances,
which we could include in our model as well as other unstable nucleon
resonances.
However, we expect that this will not change the qualitative features of the
EOS, since it depends only on the {\em total} baryon number and isospin.

Although our EOS can be calibrated very accurately at zero temperature and
normal nuclear densities, the extrapolation into the regime of high densities
is not unique.
In the context of neutron star calculations it has recently been demonstrated
\cite{HMUE96} that the high density EOS is sensitive to combinations of
parameters that
are difficult to calibrate at normal densities. As a consequence, large 
uncertainties in the predictions arise.
The density regime where the transition of hadronic
matter to a quark--gluon plasma takes place corresponds to the interior region
of a neutron star. Therefore similar difficulties can be expected.
Although this limits the quantitative results of our analysis, the
qualitative features of the basic physics remain unchanged.

Studies of the deconfinement phase transition 
in terms of a separate description of the two phases have a long history
in the literature. In particular, in the context of neutron star matter,
many calculations similar to ours have been performed earlier
\cite{G92,COLLINS75,BAYM76,KUTI80,SW86,PCL95,BDS87}.
The first analysis which considered the transition in a system with two
conserved charges was performed by Glendenning \cite{G92}, with the
restriction to zero temperature and to neutron star matter, {\em i.e.},
matter in beta-decay equilibrium.
Along these lines the effect of neutrino trapping on the composition of
cold protoneutron stars with quark cores has been investigated more recently
\cite{PCL95}.
We will extend the analysis to systems at arbitrary temperature, baryon
density and isospin. 
By definition, the two-phase model cannot reveal the actual nature of the
transition. 
In contrast, other phenomenological approaches which are directly inspired
by the symmetries of QCD, for example linear sigma models \cite{PISARSKI84},
have been applied to shed more light on the fundamental properties of the 
transition.
However, whether it is of first or second order or a cross-over
phenomenon can only be established by full QCD studies, {\em i.e.},
by lattice calculations. We believe that, until this information is
available, our approach is useful for providing a first orientation
of the qualitative features which arise in the phase diagram of strongly
interacting matter at finite baryon density and isospin.
At zero baryon density the two-phase model implies a first-order transition.
Starting from there our results are qualitatively similar to what was found in
case of the liquid--gas phase transition of nuclear matter \cite{HMUE95}. 
Primarily, the transition of hadrons to quarks and gluons in asymmetric systems
is of {\em second order}, in contrast to the first-order transition that occurs in symmetric matter (which behaves like a one-component system).
That means that some thermodynamic variables, {\em e.g.}, the entropy, are
continuous and smooth throughout the transition, whereas the first-order
transition is characterized by discontinuities in these quantities.
Furthermore, we find that the transition regime in terms of densities and
temperatures varies significantly with the isospin of the system.
In particular the density at the onset of the transition is always lower
in asymmetric matter. This result led us to the conclusion
that it is more likely that the transition region is reached in collisions
of very neutron-rich nuclei.

The outline of this paper is as follows:
In Sec.~II, we briefly summarize the general discussion 
of phase transitions in multicomponent systems which can be found in
greater detail in Ref.~\cite{HMUE95}.
We focus on the new aspects which arise when a thermodynamic system is 
described by two different equations of state.
In Sec.~III, we describe the models for the two phases
and summarize the relations that determine the equation of state. 
In Sec.~IV, we apply our model to the deconfinement phase
transition.
We study the phase coexistence region and illustrate different thermodynamic
processes by the aid of various phase diagrams and Maxwell constructions.
We consider special situations which arise during ultra-relativistic
heavy-ion collisions.
We also discuss the parameter dependence of the thermodynamic features.
Sec.~V contains a short summary.
%
%
\section{Phase Transitions in Multicomponent Systems}
We consider a system characterized by a set of $n$ mutually commuting
charges $Q_i$.
The thermodynamic features of such a system have been discussed earlier
\cite{BB80,HMUE95} in connection with the liquid-gas phase transition
of asymmetric nuclear matter.
Here we will focus on the situation where the system can exist in two
phases each characterized by a {\sl different} equation of state.
The partition function might be generated by a Hamiltonian which is
formulated in terms of different degrees of freedom in each phase \cite{G92}.
For example, the elementary constituents (quarks and gluons) would be used
in the first phase and composite particles (hadrons) in the second.
As mentioned in the introduction, it is important to keep in mind that
the conserved {\em charges\/} are the relevant quantities in the thermodynamic
analysis.
A conserved charge does not necessarily imply an independent particle
species which may change during a process (by decay, for example);
it includes any conserved quantity resulting from the
underlying symmetries of the system, for example, isospin, baryon number,
electric charge, {\em etc.}

To describe the equilibrium state of the system in phase $\alpha$
($\alpha=1,2$) enclosed in a volume $V$ we choose the
Helmholtz free energy $F$
\begin{eqnarray}
F_{\alpha}(T,V,Q_i) \equiv V{\cal F}_{\alpha}(T,\rho_i)
\quad,\quad \alpha=1,2
\label{eq:f1}
\end{eqnarray}
with
\begin{eqnarray}
\rho_i= {Q_i\over V} 
\quad,\quad i=1,\ldots , n \ .
\nonumber
\end{eqnarray}

Depending on the value of the free energy the system will be realized
in one phase or the other or in a mixture of both, indicating a phase
transition.
To be more precise, the system in phase $\alpha$ will be stable against
separation into two phases
if the free energy of the single phase $\alpha$ is lower than the free energy
in all two-phase configurations.
This requirement can be formulated as 
\begin{eqnarray}
{\cal F}_{\alpha}(T,\rho_i)<(1-\lambda){\cal F}_{\alpha}(T,\rho_i^{\alpha})
        +\lambda{\cal F}_{\beta}(T,\rho_i^{\beta})
	\quad , \quad \alpha,\beta=1,2
        \label{eq:stab1}\ ,
\end{eqnarray}
with
\begin{eqnarray}
\rho_i=(1-\lambda)\rho_i^{\alpha} +\lambda \rho_i^{\beta}\ ,
\qquad 0 < \lambda < 1\ .\label{eq:c1}
\end{eqnarray}
Note that this also implies
\begin{eqnarray}
{\cal F}_{\alpha}(T,\rho_i) \leq {\cal F}_{\beta}(T,\rho_i)
        \label{eq:stab2}\ ,
\end{eqnarray}
so if Eq.~(\ref{eq:stab1}) is satisfied, phase $\alpha$ is not only stable
against phase separation, it is also the energetically favorable single-phase
configuration.
We assume that each single phase by itself describes a stable
configuration\footnote{ When it actually happens that the free energy of
one of the single phases describes an unstable configuration, {\em e.g.}, due
to the liquid-gas phase transition in nuclear matter, we take
${\cal F}_{\alpha}$ to be the free energy resulting from a Maxwell
construction in the unstable region, which is convex.},
so that Eq.~(\ref{eq:stab1}) is satisfied for $\alpha=\beta$, {\em i.e.}, 
the free energy in each phase is a convex function of the densities \cite{RV73}.
The last equation (\ref{eq:c1}) ensures that the
overall charges are conserved:
\begin{eqnarray}
V\rho_i= V^\alpha \rho_i^\alpha + V^\beta \rho_i^\beta \quad
\hbox{with} \qquad V&=& V^\alpha + V^\beta \ .\label{eq:c2}
\end{eqnarray}

Equation (\ref{eq:stab1}) is a {\em global\/} criterion for
the stability of a one-phase configuration.
Whenever it is violated, a system with 
two phases is energetically favorable. 
The phase coexistence is governed by the Gibbs' conditions \cite{LL59}
demanding equal pressure $p$, temperature $T$ and chemical potentials $\mu_i$
in both phases
\begin{eqnarray}
\mu_i^\alpha (T,\rho_i^\alpha)&=&\mu_i^\beta(T,\rho_i^\beta) \ ,
\label{eq:g1}\\
p^\alpha(T,\rho_i^\alpha)&=&p^\beta(T,\rho_i^\beta) \ .\label{eq:g2}
\end{eqnarray}
It is important to realize, however, that under conditions of phase 
separation, Eq.~(\ref{eq:stab1}) may still be valid {\em locally\/},
that means, in some small region $U=\{T,\rho_i\}$ of parameter space,
but it may nevertheless be possible 
to find significantly different densities $\rho_i^{\alpha}$ and 
$\rho_i^{\beta}$ in a larger domain $ U\subset G$ that violate this condition.
This leads to the existence of metastable states.

The two sets of densities $\{\rho_i^{\alpha},\rho_i^{\beta}\}$
that satisfy Eqs.~(\ref{eq:g1}) and (\ref{eq:g2}) form a surface
in the parameter space $\{T,\rho_i\}$; this is the phase separation boundary,
or binodal. 
For $n$ conserved charges and two coexisting phases, Gibbs' phase rule implies
that the binodal is an $n$-dimensional surface \cite{LL59}.
It can also be shown that this surface encloses all points 
that lead to a single (unstable) configuration with a higher value
for the free energy \cite{HMUE95}. 

The binodal surface determines the stability boundaries of the system.
The mixed phase inside must be determined by a Maxwell construction.
This is achieved by solving Eq.~(\ref{eq:c1}) \cite{BB80,HMUE95}
%
%
for given values of $\rho_i$,
with $\rho_i^{\alpha}$ and $\rho_i^{\beta}$ lying on the binodal surface.
The free energy in the transition region is then given by
\begin{eqnarray}
{\cal F}(T,\rho_i)=(1-\lambda){\cal F}_{\alpha} (T,\rho_i^{\alpha})
+\lambda{\cal F}_{\beta}(T,\rho_i^{\beta})\ . \label{eq:femax}
\end{eqnarray}
Densities related to other extensive quantities
can be computed accordingly.

These ideas are illustrated in Fig.~(\ref{fig:max}).
We show the free energy density in both phases as a function of the density
coordinates $\{\rho_i\}$ at constant temperature. At the location of the points
$B_1$ and $B_2$, the system enters the binodal region. In between, a phase
mixture is energetically favorable. The free energy of the mixed phase as it
results from the Maxwell construction Eq.~(\ref{eq:femax}) is indicated by
the solid line which connects $B_1$ with $B_2$. The dotted lines
correspond to the free energy of the single-phase configurations. Between
the point of intersection $I$ and $B_1$ and between $I$ and $B_2$ lie
the metastable states of phase 1 and phase 2, respectively. The configurations
beyond $I$ are unstable. Here not only a phase mixture but also the single
(metastable) configuration of the opposite phase is energetically favorable.
Instability normally arises due to fluctuations, which no longer can be
restored by the system. This is signaled by the violation of a local
stability criterion, {\em e.g.}, a negative compressibility.
However, our description is based purely on the energetics and cannot
provide information about the physical mechanism which leads to instability.
This is certainly an indication that the separate description of the
two phases breaks down at this point.

We close this section by specializing the general formalism to asymmetric
matter.
In the hadronic phase it consists of strongly interacting nucleons and mesons,
and in the deconfined phase, weakly interacting quarks and gluons.
For two quark flavors
such a system is characterized by two conserved charges: 
the total baryon number 
\begin{eqnarray}
N_B \equiv V \rho_B 
\end{eqnarray}
and the third component of isospin
\begin{eqnarray}
I_3 \equiv {1\over 2}\, V\rho_3 \ .
\end{eqnarray}
Thus we have for the thermodynamic potential, or equivalently the pressure,
\begin{eqnarray}
{\Omega_\alpha \over V}(T,\mu_B^\alpha,\mu_3^\alpha)= -p^\alpha =
{\cal F}_\alpha(T,\rho_B,\rho_3)
      -\mu_B^\alpha\rho_B-{1\over 2}\mu_3^\alpha\rho_3\ ,
\end{eqnarray}
where the baryon and isospin chemical potentials are given by
\begin{eqnarray}
\mu_B^\alpha =
   \left({\partial{\cal F}_\alpha
    \over \partial \rho_B}\right)_{T,\rho_3}
\quad , \quad
\mu_3^\alpha =
     2\left({\partial{\cal F}_\alpha
     \over \partial \rho_3}\right)_{T,\rho_B} \ .
\label{eq:chpot}
\end{eqnarray}
In the discussion of asymmetric systems it is also useful to introduce the 
isospin ratio
\begin{eqnarray}
x \equiv {I_3 \over N_B}={\rho_3 \over 2\rho_B} \ ,
\label{eq:defx}
\end{eqnarray}
so that the free energy can be rewritten as
\begin{eqnarray}
{\cal F}_\alpha(T,\rho_B,\rho_3)=
       {\cal F}_\alpha(T,\rho_B,x) \ .
\label{eq:fx}
\end{eqnarray}
%
%
\section{The Equation of State}
To generate the nuclear equation of state in the hadronic phase, we adopt
the relativistic mean-field approach of Refs.~\cite{FST96,HMUE96}
involving valence Dirac nucleons and effective mesonic degrees of freedom, 
which are taken to be neutral scalar $(\phi) $ and vector fields $(V^{\mu})$,
plus the isovector $\rho$ meson $({\bf b}^{\mu})$ and pion field
$(\bbox{\pi})$.
The basic quantity here is the thermodynamic potential as a functional of
the meson mean fields which, in principle, can be formulated without reference
to an explicit lagrangian. 
For homogeneous nuclear matter, at finite temperature and chemical potentials
$(\mu_B,\mu_3)$, it is of the general form \cite{HMUE96}
\begin{eqnarray}
 {\Omega\over V}(T, \mu_B, \mu_3;\phi,V_{\nu},{\bf b}_{\nu},{\boldpi})&=&
               {\Omega_{N}\over V}+ {\Omega_{M}\over V}
 + {1\over 2}\,m_{\rm s}^2 \phi^2
      -{1\over 2}\,m_{\rm v}^2 V_{\mu} V^{\mu}
          -{1 \over 2}\, m_{\rho}^2{\bf b_{\mu}}
                             \bolddot{\bf b}^{\mu}
 + {1\over 2}\,m_{\pi}^2 {\boldpi}^2 \nonumber\\
& &\null + \Delta{\cal V}(T, \mu_B, \mu_3; \phi,V_{\mu},
            {\bf b_{\nu}},{\boldpi})\ .
			   \label{eq:Omh}
                      \end{eqnarray}
The first two terms in Eq.~(\ref{eq:Omh}), $\Omega_{N}/V$ and $\Omega_{M}/V$,
are the one-body contributions due to valence nucleons and mesons. 
The nonlinear potential $\Delta{\cal V}$ represents the unknown part of
the thermodynamic potential, which includes the effects of nucleon 
exchange and correlations,
as well as contributions from the quantum vacuum \cite{FST96,FTS95}.

Following Ref.~\cite{HMUE95} the one-body fermionic contribution can be written
as 
\begin{eqnarray}
{\Omega_{N}\over V} =
   -{1\over 3\pi^2}\Bigl[H_5(\nu_p,M^*)+H_5(\nu_n,M^*)\Bigr] \ ,
\label{eq:Omhn}
\end{eqnarray}
where the integral $H_5$ as well as others will be defined shortly.
The corresponding contributions of the heavy meson fields
$(\phi,V^{\mu},{\bf b}^{\mu})$ to $\Omega_{M}$ are negligible
in the relevant temperature range $(T\lesssim 200$ MeV).
Thus we keep only the pions and write
\begin{eqnarray}
{\Omega_{M}\over V} = 
{\Omega_{\pi}\over V} = -{1\over 6\pi^2}\Bigl[B_5(\nu_{\pi},m_{\pi})
                              + {1\over 2} B_5(0,m_{\pi})\Bigr] \ ,
\label{eq:Omhm}
\end{eqnarray}
where the first term is the contribution of the charged pions
\cite{HABER82,KAPUSTA89}, and the latter arises from the $\pi^0$. 
The nonlinear potential $\Delta{\cal V}$ can be expanded in a Taylor series 
in terms of the meson mean fields \cite{FST96,HMUE96}. 
In the normal phase of nuclear matter, {\em i.e.}, without a pion condensate,
the pion mean field vanishes and the pions contribute only to the coefficients 
of this series via loops.
In practice, the series must be truncated, and the unknown coefficients
serve as model parameters, which are chosen to reproduce certain empirical
properties of equilibrium nuclear matter, as discussed below.
For our calculation we choose the explicit form
\begin{eqnarray}
\Delta {\cal V}= {\kappa\over 3!}\phi^3+{\lambda\over 4!}\phi^4
		-{\zeta\over 4!}g_{\rm v}^4(V_{\mu}V^{\mu})^2 \ ,
					  \label{eq:pot}
\end{eqnarray}
which includes a subset of the
meson self-interactions up to fourth order in the fields.
In principle, terms which couple different fields
and also nonlinear interactions involving the $\rho$ meson field 
are allowed \cite{FST96}.
We also disregard contributions involving the
pion mean field which arise in the pion condensed phase.
Although disregarding these couplings  is ``unnatural'',  the present model
defined by Eq.~(\ref{eq:pot}) is already general enough for our purposes.
First of all, it can be accurately calibrated at normal densities, and
it can be related to the most common models discussed in the literature.
Unless one assumes scalar self-interactions only, the nonlinear couplings
cannot be uniquely constrained by the
calibration procedure, leading to uncertainties in the high-density EOS.
In particular, little information is available to 
constrain contributions which explicitly depend on the pion field.
Moreover, among these nonlinear terms, the quartic vector coupling in
Eq.~(\ref{eq:pot}) is the most important one at high densities \cite{HMUE96}.
We will use this coupling to examine the uncertainties which arise in the
quantitative predictions for the transition from hadrons to a
quark-gluon plasma.

The mean fields are determined by extremization of the thermodynamic
potential Eq.~(\ref{eq:Omh}).
The specific form of $\Delta{\cal V}$, introduced in Eq.~(\ref{eq:pot}),
together with Eqs.~(\ref{eq:Omhn}) and (\ref{eq:Omhm}), leads to the
self-consistency equations

\begin{eqnarray}
{m_{\rm s}^2 \over g_{\rm s}^2}\, \Phi
  +{\kappa \over 2 g_{\rm s}^3}\, \Phi^2
  +{\lambda \over 6 g_{\rm s}^4}\, \Phi^3 &=& \rho_s \ ,\label{eq:scalar}\\
W \left( 1 + {g_{\rm v}^2 \over m_{\rm v}^2}\,{\zeta\over 6}\, W^2\right)
        &=& {g_{\rm v}^2 \over m_{\rm v}^2}\, \rho\ , \label{eq:vector}\\
R &=& {g_{\rho}^2 \over 2 m_{\rho}^2}\, \rho_3 \ , \label{eq:Rrho}
\end{eqnarray}
where the scalar and baryon densities are given by
\begin{eqnarray}
\rho_s &=& {M^*\over \pi^2}\Bigl[H_3(\nu_p,M^*)+H_3(\nu_n,M^*)\Bigr] \ ,
           \label{eq:rhos}\\
\rho_B &=& {1\over \pi^2}\Bigl[G_3(\nu_p,M^*)+G_3(\nu_n,M^*)\Bigr] \ .
           \label{eq:rhob}
\end{eqnarray}
The isospin density receives contributions from nucleons and pions
\begin{eqnarray}
{1\over 2}\rho_3 &=& {1\over 2}\rho_3^N + {1\over 2}\rho_3^{\pi}
           \label{eq:rho3}\\
&\equiv& {1\over 2\pi^2}\Bigl[G_3(\nu_p,M^*)-G_3(\nu_n,M^*)] 
            +{1\over 2\pi^2}A_3(\nu_{\pi},m_{\pi}) \ .
           \nonumber
\end{eqnarray}
The scaled meson fields are 
$\Phi \equiv g_{\rm s} \phi$, $W \equiv g_{\rm v} V_0$, and
$R \equiv g_\rho b_0$, with $b_0$ the timelike, neutral part of the $\rho$
meson field.

The baryon effective mass and effective chemical potentials are defined in
terms of the meson mean fields and the baryon and isospin chemical potential
as
\begin{eqnarray}
M^* &\equiv& M- \Phi \ , \label{eq:mstar}\\
\nu_p &\equiv& \mu_B+{\mu_3\over 2} - W - {1\over 2}\, R \ , \label{eq:nup}\\
\nu_n &\equiv& \mu_B -{\mu_3\over 2} - W + {1\over 2}\, R \ . \label{eq:nun}
\end{eqnarray}
The effective chemical potential for the pions is given by
\begin{eqnarray}
\nu_{\pi} \equiv \nu_p-\nu_n = \mu_3- R \ , \label{eq:nupi}
\end{eqnarray}
so that the $\rho$ meson field in Eq.~(\ref{eq:Rrho}) couples to the total
isospin density in Eq.~(\ref{eq:rho3}).
We also introduce the required integrals over the thermal distribution
functions as
\begin{eqnarray}
G_{n}(\mu,M)&\equiv&\int_{0}^{\infty} k^{n-1} dk
\left( {1\over 1+{\rm e}^{\mkern2mu\beta [E(k,M) - \mu ]}}
      - {1\over 1+{\rm e}^{\mkern2mu\beta [E(k,M) + \mu ]}}\right) \ , 
           \label{eq:Gdef}\\[4pt]
H_{n}(\mu,M)&\equiv&\int_{0}^{\infty} {k^{n-1} dk \over E(k,M)} 
  \left( {1\over 1 + {\rm e}^{\mkern2mu\beta [E(k,M) - \mu ]}}
      + {1\over 1 + {\rm e}^{\mkern2mu\beta [E(k,M) + \mu ]}}\right)\ , 
           \label{eq:Hdef}
\end{eqnarray}
for fermions and
\begin{eqnarray}
A_{n}(\mu,m)&\equiv&\int_{0}^{\infty} k^{n-1} dk
\left( {1\over {\rm e}^{\mkern2mu\beta [E(k,m) - \mu ]}-1}
      - {1\over {\rm e}^{\mkern2mu\beta [E(k,m) + \mu ]}-1}\right) \ , 
           \label{eq:Adef}\\[4pt]
B_{n}(\mu,m)&\equiv&\int_{0}^{\infty} {k^{n-1} dk \over E(k,m)} 
  \left( {1\over {\rm e}^{\mkern2mu\beta [E(k,m) - \mu ]}-1}
      + {1\over  {\rm e}^{\mkern2mu\beta [E(k,m) + \mu ]}-1}\right)\ , 
           \label{eq:Bdef}
\end{eqnarray}
for bosons,
where $E(k,M) \equiv (k^2+M^2)^{1/2}$, and $n > 0$ to ensure convergence is 
understood. Moreover, the boson integrals are subject to the constraint
$|\mu|\leq m$.

By inserting the solution of
Eqs.~(\ref{eq:scalar})--(\ref{eq:Rrho}) into Eq.~(\ref{eq:Omh}), 
it is straightforward to compute the pressure
\begin{eqnarray} p = -{\Omega\over V}&=&
     -{\Omega_{N}\over V}(T,\nu_p,\nu_n, M^*)
     -{\Omega_{\pi}\over V}(T,\nu_{\pi})
      \nonumber\\
 & &\null + {m_{\rm v}^2\over 2g_{\rm v}^2}\, W^2 + {\zeta\over 24}\,W^4
          +{m_{\rho}^2\over 2g_{\rho}^2}\, R^2 
          -{m_{\rm s}^2\over 2g_{\rm s}^2}\, \Phi^2
          -{\kappa\over 6 g_{\rm s}^3 }\Phi^3
	  -{\lambda\over 24 g_{\rm s}^4 }\Phi^4 \ .
           \label{eq:pressh}
\end{eqnarray}
Other quantities like the free energy, the entropy, {\sl etc.}, follow by using
standard thermodynamic relations.

For the quark-gluon phase we adopt a bag-model type EOS \cite{KUTI80}
involving massless $u$ and $d$ quarks $(N_f=2)$
\begin{eqnarray}
{\Omega\over V}(T,\mu_B,\mu_3) =
     {\Omega_{\rm{\scriptscriptstyle pert}}\over V}(T,\mu_B,\mu_3) + b \ ,
\label{eq:Omq}
\end{eqnarray}
where $\Omega_{\rm{\scriptscriptstyle pert}}/ V$ is the perturbative expansion
of the thermodynamic potential 
\begin{eqnarray}
{\Omega_{\rm{\scriptscriptstyle pert}}\over V}(T,\mu_B,\mu_3)&=&
-{\pi^2\over 45} T^4 \Bigl(8+{21\over 4}N_f\Bigr)
-{1\over 2}\sum_{f=u,d}\Bigl( T^2 \mu_f^2 +{\mu_f^4\over 2\pi^2}\Bigr)
\nonumber\\
& &\null + {2 \pi\over 9}\alpha_s
\biggl(T^4 \Bigl[3+{5\over 4}N_f\Bigr]
+{9\over 2}\sum_{f=u,d}\Bigl[ {T^2 \mu_f^2\over \pi^2}
			     +{\mu_f^4\over 2 \pi^4}\Bigr] \biggr) \ ,
\label{eq:Omqp}
\end{eqnarray}
and $b$ is the difference between
the energy density of the perturbative and the nonperturbative QCD vacuum, 
{\em i.e.}, the bag constant.
The chemical potentials of the two flavors are connected to the baryon and
isospin chemical potentials by
\begin{eqnarray}
\mu_u = {1\over 3}\mu_B+{1\over 2}\mu_3
\quad , \quad
\mu_d = {1\over 3}\mu_B-{1\over 2}\mu_3 \ ,
\label{eq:muq}
\end{eqnarray}
which allows the computation of 
the corresponding densities according to
\begin{eqnarray}
\rho_B = -\left( {\partial\over\partial\mu_B}{\Omega\over V}\right)_{T,\mu_3}
\quad , \quad
{1\over 2} \rho_3 = -\left( {\partial\over\partial\mu_3}
{\Omega\over V}\right)_{T,\mu_B} \ .
\label{eq:rhoq}
\end{eqnarray}

We close this section with a short description of the calibration of the EOS.
The model for the hadronic phase has six free parameters.
Following Ref.~\cite{FST96} we choose the value of the coupling $\zeta$ and
determine the other parameters so that the five equilibrium properties
of nuclear matter,
as listed in Table~\ref{tab:one}, are reproduced.
The free parameter $\zeta$ is chosen within the natural range 
$0\leq\zeta\leq0.06$ \cite{FST96,HMUE96}.
To fix the two parameters in the quark-gluon phase, namely the
bag constant $b$ and the strong coupling constant $\alpha_s$, we proceed
as follows. We specify the bag constant to $b = 160$ MeV/fm$^3$ and determine
$\alpha_s$ so that our model reproduces a transition temperature of 150 MeV
at zero baryon density, which is in the currently accepted temperature range
\cite{BLUM95}. As a consequence of this calibration $\alpha_s$ depends on
$\zeta$. 
This dependence, however, is extremely weak. The value $\alpha_s= 0.349$
obtained for $\zeta=0$ changes only by $0.1\%$ when $\zeta$ is 
varied within the natural range.
%
%
\section{Phase structure of asymmetric matter}
In this section we apply the formalism of Sec.~II to the
deconfinement phase transition.
Within our simple two-phase model
we consider highly excited asymmetric matter created in
an ultra-relativistic collision of two heavy-ions.
The dynamical evolution of the system during the collision is a very complex
process and it is convenient to divide the reaction into three stages:
the formation of a highly excited matter system, its expansion and the eventual
decay.
The question of whether the required energy densities can be achieved in an
experiment to form a 
quark--gluon plasma during the first violent stage is a difficult one, which
we will not attempt to answer here.
We simply assume that such a state arises and 
ultimately reaches equilibrium at some finite temperature, density and pressure.
We follow the subsequent expansion of the system during the second stage and
concentrate on the new features of the phase diagram that arise as a
function of the total isospin.
Although our general discussion will encompass the entire range of $x\leq 0$,
we will study systems with $-0.2\leq x\leq 0$ and $x=-0.5$ in more detail
to obtain estimates for the size of the new effects. The former range 
covers the isospin ratios which are experimentally accessible in
heavy--ion collisions,
{\sl e.g.}, $x_{{\rm Au}+{\rm Au}},x_{{\rm U}+{\rm U}}\approx -0.1$,
whereas the latter value corresponds to neutron matter relevant in
astrophysical applications.

As discussed in the context of the liquid-gas phase transition of asymmetric
nuclear matter \cite{BB80,HMUE95}, the phase structure of binary systems
is more complex than in one-component systems. 
The basic quantity in the analysis is the phase separation boundary or binodal,
which forms a two-dimensional surface in parameter space
as discussed in Sec.~II.
Before we discuss the binodal structure of the deconfinement
phase transition as it arises in our model, let us briefly point out
the role of the pions.
To this end, assume our system is prepared in the hadron phase at some fixed
isospin ratio $x$, as defined in Eq.~(\ref{eq:defx}), and let us follow an
isothermal compression. At some critical baryon density $\rho_B^c$ the chemical
potential of the pions will approach the (negative) value of the pion mass,
{\em i.e.},
\begin{eqnarray}
\nu_{\pi} \to -m_{\pi}
\quad \hbox{for} \quad
\rho_B \to \rho_B^c \ , \label{eq:cond}
\end{eqnarray}
indicating the onset of a pion condensate \cite{LL59}. The critical density
depends on $x$ and on $T$, and will be smaller in more asymmetric systems.
At this point a macroscopically occupied $\pi^-$ mode arises which
is characterized by a nonvanishing pion mean field. In principle, the 
corresponding contributions to the thermodynamic potential in Eq.~(\ref{eq:Omh})
must now be kept explicitly, and the value of the mean field is determined
by minimization \cite{HABER82,BAYM73}. 
As discussed in the last section we will disregard the explicit
mean-field contributions of the pions in the following.
That means we neglect the pion-nucleon as well as the pion self-interactions.
To incorporate the condensed phase consistently we will be satisfied 
to add the contribution of the zero-momentum states of the pions to the
isospin density in Eq.~(\ref{eq:rho3})
\begin{eqnarray}
{1\over 2}\rho_3^{\pi}=-2 m_{\pi}\kappa^2
		   + {1\over 2\pi^2}A_3(-m_{\pi},m_{\pi}) \ ,
		                   \label{eq:rho3PiC}
\end{eqnarray}
where the effective chemical potential 
\begin{eqnarray}
\nu_{\pi}=\nu_p-\nu_n= -m_{\pi}
\label{eq:nuPiC}
\end{eqnarray}
is held fixed in the condensed phase. The parameter $\kappa^2$ 
is the continuous order parameter of this transition, which
specifies the amount of isospin charge carried by the zero-momentum states
\cite{HABER82,KAPUSTA89}.
In contrast, the condensate does not contribute to the pressure
of the system; thus, one simply substitutes the value of $\nu_{\pi}$, as given
by Eq.~(\ref{eq:nuPiC}), in the expression for the hadronic
pressure Eq.(\ref{eq:pressh}). We will leave the study of the contributions
arising from pionic interactions as an important topic for future work.

We now return to the general discussion of the phase diagram in
our system.
We begin the analysis with $\zeta=0$ for the value of the quartic vector meson
coupling in the hadronic EOS. We will study the influence of this parameter
at the end of this section.

The binodal is determined by Gibbs' conditions 
(\ref{eq:g1}) and (\ref{eq:g2}) which enforce
equal pressure and chemical potentials for the hadronic phase and for the 
quark--gluon phase in equilibrium.
In general these two phases have a different isospin ratio $x$.
The binodal surface in $\{x,p,T\}$ space is indicated in Fig.~\ref{fig:slices}. 
A sequence of several slices at fixed $T$ is shown.
For a given temperature,
the binodal section is divided into two branches.
One branch, at the lower pressure, describes the system in the hadron phase,
while the other branch, at the higher pressure, describes the quark--gluon
phase.
These two branches contain the beginning and ending configurations of the
phase transition.
The shape of the binodal slices changes drastically
between $T=0$ and $T=150$ MeV.
One observes that the enclosed area becomes smaller with increasing
temperature until, at $T_c = 150$ MeV,
the two branches coalesce to a single line at constant pressure $p = 19.08$
MeV$/$fm$^3$ with isospin ratios ranging from $x\approx-4.5$ to $x=0$.
This line of critical points marks the transition point of systems at
zero baryon density.

Another important observation is that the pressure steadily decreases
when the temperature increases.
At the same time, the transition regime, {\em i.e.}, the difference in pressure
between the two phases at the beginning and at the end of a transition, becomes
smaller.
Also indicated in Fig.~(\ref{fig:slices}) are the special
configurations that separate into two phases with the same isospin ratio.
The coordinates of these configurations form the two lines of equal
concentration (LEC), one of which coincides with the projection $x= 0 $ in our
system. 
The isospin ratio of the second LEC increases with temperature and
eventually the two LEC intersect at $T\approx 137.2$ MeV.
For a fixed temperature the pressure necessarily attains an extremum
at the points of equal concentration \cite{LL59}.

Generally, the hadronic configurations are divided into normal states and
states exhibiting
a pion condensate. The coordinates of the configurations at the onset
of the condensation also form a two--dimensional surface. This critical
surface can be parametrized in terms of two coordinates in $\{x,p,T\}$ space,
{\sl e.g.}, as $p_c(x,T)$.
The intersection of the critical surface with the binodal forms the
line of three phases (LTP),
which marks the onset of the pion condensate on the hadronic 
branch\footnote{Since there is no two-phase coexistence
in a Bose condensation,
the LTP is not a line of triple points which could arise 
in a binary system \cite{LL59}.  The LTP is merely a boundary between areas
in parameter space which describe two different two-phase configurations.}.
In Fig.~(\ref{fig:slices}) these states are emphasized by the bold dashed
lines. The isospin ratio of the LTP decreases with increasing temperature. Above
$T\approx 136.2$ MeV the critical pressure curve and the binodal no longer
intersect.

The binodal in $\{\rho_B,x,T\}$ space is indicated in
Fig.~(\ref{fig:bintvsrho}),
where several branches at different $x$ are projected onto the
$\{\rho_B,T\}$ plane. The onset of the condensate,
{\em i.e.}, the critical temperature as a function of the baryon density and 
isospin ratio, is marked by an open circle. Note the change of density
in the two--phase region. The most extreme situations are encountered at $T=0$.
For example, in symmetric matter the density increases between the onset and
the completion of the transition by more than 3.5$\rho_B^0$.
This density regime becomes even larger in asymmetric systems.
Another important observation is that the density at the onset decreases
with increasing asymmetry. If we compare these points for $x=0$ and $x=-0.2$
at $T=0$ we find an decrease of $\Delta\rho_B\approx 0.5\rho_B^0$.

The role of the pions becomes more apparent in Fig.~(\ref{fig:slice}).
Part (a) indicates one slice of Fig.~(\ref{fig:slices}) at $T = 50$ MeV.
Also included is the projection of the critical surface in the form of the
critical pressure curve $p_c (x)$.
Nuclear matter in the normal phase resides on the right-hand side,
and the states with a pion condensate on the left-hand side.
The critical curve intersects the binodal at TP.
In contrast, the pion contributions are not included in part (b). With nucleons
only, the isospin ratio in the hadronic branch is limited by
$-0.5 \leq x \leq 0$ which leads to a widely open binodal surface.
Furthermore, only one point of equal concentration arises (at $x = 0$).

To illustrate the phase-separation scenarios,
we study the behavior of matter under an isothermal compression.
Assume that the system is initially prepared in
the hadronic phase with isospin ratio $x = -0.4$ and $p\lesssim 50$ MeV/fm$^3$.
The situation is also indicated in part (a) of Fig.~\ref{fig:slice}.
During the compression, the system first crosses the critical pressure
curve at the point $A$, where pion condensation begins.
Starting with $\kappa^2=0$ at $A$ the contribution of the condensate to
the (fixed) isospin ratio increases during the compression.
At the point $B$ the two-phase region is encountered and now a quark--gluon
phase is about to emerge. 
The solution of Gibbs' conditions
(\ref{eq:g1}) and (\ref{eq:g2})
determine the density and the isospin ratio $x_{\scriptscriptstyle C}$ of this
new phase, which occurs at the point labeled $C$.
As the system is compressed, the {\em total\/} isospin ratio $x$ remains
fixed, as dictated by the conservation laws, but the hadronic phase evolves 
from $B$ to $E$, while the quark--gluon phase evolves from $C$ to $D$.
At the point $D$, the system leaves the region of instability.
The original hadronic phase (which now has no pion condensate) is about to
disappear, and it exists in an 
infinitesimal volume with a density and isospin ratio
$x_{\scriptscriptstyle E}$ corresponding to the point $E$.

An important observation is that if the system encounters the binodal
between the two points of equal concentration then the quark--gluon phase is
always more asymmetric and has a higher baryon density than the hadronic
phase.
This behavior is due to the {\sl symmetry energy}, which is higher in the
hadronic phase. Therefore it is energetically
favorable for asymmetric matter to separate into a denser and
more asymmetric quark--gluon phase and a hadronic phase that is more
dilute and symmetric.

The energetics also implies a significant simplification if the
system is initially prepared with an isospin ratio corresponding
to one of the points EC.
It is now energetically favorable for the system to separate into two phases
with the same composition. In this so-called ``indifferent
equilibrium'' \cite{HS69}, matter behaves like a 
one--component system (see also the discussion in \cite{HMUE95}). In contrast 
to the previous case the system becomes unstable at one of the points EC in 
Fig.~\ref{fig:slice} 
and stays at this point until the transition is completed.
Note that the location of the point EC at $x=0$ is model independent
since symmetric matter is always the configuration with lowest energy.
The location of the second point depends on specific model features,
particularly on the predicted values for the symmetry energy and on the
pionic contributions.

The results of the corresponding Maxwell constructions are indicated
in Fig.~\ref{fig:pvsrho}.
We consider different isospin ratios at $T=0$ in part (a) and at $T= 50$ MeV
in part (b), respectively.
For symmetric matter $(x = 0)$, we obtain the familiar result
with a constant vapor pressure, represented by a horizontal line.
At $x<0$ the equal demand of Gibbs' conditions and the conservation of the
overall isospin forces the pressure to change during the transition.
Also shown in Fig.~\ref{fig:pvsrho} are isotherms for ``neutron matter''
$(x=-0.5)$. The name is somewhat misleading since the system contains
a finite number of protons even without the pions.
It is common in the literature \cite{HMUE96,COLLINS75,BAYM76,BDS87} to treat
neutron matter as a single-component system.
The qualitatively different features which
arise in a proper treatment as a system with {\sl two} conserved charges
have been pointed out earlier by Glendenning \cite{G92} in the context
of neutron star calculations (see also Ref.~\cite{PCL95}).
In addition, part (b) of Fig.~\ref{fig:pvsrho} also includes a system with an
isospin ratio which corresponds to the second point of equal concentration.
As discussed above, this system also stays at a constant vapor 
pressure throughout the transition.

The change of the pressure throughout the phase separation
in asymmetric systems is an indication of
a {\sl smoother} transition than in symmetric or one--component systems.
A more precise characterization of the transition can be obtained from
the isobaric process indicated in Fig.~(\ref{fig:svst}). 
The entropy as a function
of temperature for three different isospin ratios 
is shown.
In symmetric matter we find a discontinuity at the transition temperature,
which gives rise to a latent heat $Q_L$, {\em i.e.}, all the heat is used to
convert hadrons into quarks.
In asymmetric matter the entropy remains
smooth but the temperature now changes throughout the transition so that no
strict latent heat can be assigned to this process.
For example, the isospin ratio $x= -0.2$ shown in Fig.~(\ref{fig:svst}) gives
rise to a temperature change of $\Delta T\approx 30$ MeV.
As discussed in more detail in Ref.~\cite{HMUE95}, these are the features
which lead to a first-order transition in symmetric matter and to a
second-order transition in the asymmetric system,
according to Ehrenfest's definition.

The processes we have considered so far correspond to very specific idealized
situations and served mainly to study the basic thermodynamic features of
our system. 
Generally, it is not clear if the expansion of the highly excited matter
created in a heavy--ion collision can be described by a distinct thermodynamic
process.
However, the situation simplifies at extremely high collision energies.
In this so-called scaling regime, the colliding nuclei become transparent and
the excited plasma is created between the two nuclei when they recede from 
the collision point. The basic assumption here is that the total entropy
is created in the initial stage of the collision and remains nearly constant
during the later evolution. Under this assumption, the expansion of the plasma
is nearly adiabatic and can be described by relativistic hydrodynamics
\cite{CHIU75,BJORK83}.

To make contact with this discussion, we also studied adiabatic processes.
The properties of the system in the $(\rho_B,T)$ plane can be studied in 
Fig.~(\ref{fig:tvsrhos}).
Adiabats for various values of the entropy per baryon are indicated for
symmetric matter in part (a) and for asymmetric matter in part (b),
respectively. The dashed lines are the coexistence curves, {\em i.e.}, the
projection of the hadronic and the quark--gluon branch of the binodal,
which intersect at the critical point CP.
The curves in the coexistence region are the
result of a Maxwell construction.
Note the behavior of the system in the mixed phase.
Throughout the transition the temperature {\em decreases}
with increasing density,
in contrast to the pure phases where the temperature increases monotonically.
This peculiar behavior arises because for a given baryon density and
temperature, the entropy in the quark--gluon phase is always higher.
If we think in terms of a ``boiling'' or ``condensation'' process and compare
to a liquid-gas transition one would expect the opposite
behavior \footnote{A system which exhibits a similar behavior is the
liquid-solid phase transition of $^3$He, for which the entropy in the
high-density solid phase is higher than in the low-density liquid phase due to
the spin disorder present in the solid phase \cite{GOLD92}.}.
In an ordinary liquid the high-density phase has the lower entropy and the
temperature would rise steadily during the transition.

If we compare the extension of the transition regime in the symmetric and
asymmetric systems, we find a similar quantitative difference as observed in
the previous discussion. Generally, for a given value of the entropy per
baryon, the onset of the transition occurs at a lower density and temperature
in asymmetric matter. For example, if we compare the adiabat $S/N_B = 2$ in
Fig.~(\ref{fig:tvsrhos}a) and (\ref{fig:tvsrhos}b), the density 
decreases by $\Delta\rho_B\approx \rho_B^0/3$ and the temperature 
by $\Delta T \approx 4$ MeV.
Compared to the overall scale set by the critical temperature of $150$ MeV
this temperature difference is certainly less significant.
However, the qualitative difference between
the two systems could be an indication that the transition region is
experimentally better accessible in a collision of extremely neutron-rich
systems, which might be obtainable with radioactive beams.

Finally, we briefly discuss how the properties of the phase transition 
change when the parameter $\zeta$ in the hadronic EOS is varied.
This point has been discussed in Ref.~\cite{HMUE96} for pure neutron matter
and here we will extend the analysis to arbitrary isospin ratios.
Up to now the hadronic EOS in Eq.~(\ref{eq:pressh}) was generated  
without the quartic vector meson interaction.
In the following we allow a nonzero coupling within the natural range 
$0\leq \zeta \leq 0.06$.

Figure~(\ref{fig:rhovszeta}) shows the baryon density at the onset and at the 
end of the transition for the isospin ratios $x=0, -0.2, -0.5$ at $T=0$.
Note that all values of $\zeta$ reproduce identical properties of nuclear
matter and a transition temperature of $T=150$ MeV at zero baryon density.
Generally, one observes that for arbitrary isospin ratios, increasing
$\zeta$ softens the EOS, which leads to substantially higher transition
densities.
The increase in the symmetric system is remarkable (note the logarithmic scale);
at sufficiently large values of $\zeta$ the transition vanishes altogether.
The transition is driven by the energetics in both phases and therefore
this behavior is most easily understood by examining the high-density limit of
the energy per baryon \cite{HMUE96}.
Using Eq.~(\ref{eq:Omqp}) and the relation ${\cal E}=3p+4b/3$,
one obtains for the quark phase
\begin{eqnarray}
{\lim_{{\rho_B}\to \infty}}
 {\cal E}_Q/ \rho_B = c_Q\rho_B^{1/3} \ .
 \label{eq:epsq}
\end{eqnarray}
The asymptotic behavior of the hadronic EOS depends on $\zeta$ and $x$.
For $x<0$, $\zeta\geq 0$ or $x=0$, $\zeta =0$ the quadratic terms
in Eq.~(\ref{eq:pressh}) dominate, which leads to
\begin{eqnarray}
{\lim_{{\rho_B}\to \infty}}
 {\cal E}_H/ \rho_B \propto \rho_B \ .
 \label{eq:epsh1}
\end{eqnarray}
Thus, at sufficiently high densities, the hadronic matter always has higher
energy compared to the quark phase and a transition is possible.

A special situation arises for $x=0$ and $\zeta>0$. In this case the
leading behavior is, up to the prefactor, identical to Eq.~(\ref{eq:epsq}),
{\em i.e.},
\begin{eqnarray}
{\lim_{{\rho_B}\to \infty}}
 {\cal E}_H/ \rho_B = c_H\rho_B^{1/3} \ .
 \label{eq:epsh2}
\end{eqnarray}
By comparing the factors $c_Q$ and $c_H$ in
Eq.~(\ref{eq:epsq}) and in Eq.~(\ref{eq:epsh2}), respectively, one arrives at
the remarable result that a transition is possible only if 
(see also the discussion in \cite{HMUE96})
\begin{eqnarray}
{1\over 3}\left[1+\Bigl({4\over \pi^2\zeta}\Bigr)^{1/3}\right] 
		  > 1+{2\alpha_s\over 3\pi}
       \label{eq:condition} \ .
\end{eqnarray}
This implies that for sufficiently large values of the nonlinear
coupling, {\em symmetric matter remains in the hadron phase}.
Note that we would arrive at this result for {\em arbitrary} isospin
ratios if we included an additional quartic $\rho$ meson coupling.
In accordance with Ref.~\cite{HMUE96}, we conclude that the parameter
dependence of the hadronic EOS at high densities leads to large uncertainties
in the predictions for the transition region of the
deconfinement phase transition (if it exists).
In principle, reproducing the quantitative features of the transition
could be a tool for calibrating the hadronic EOS better in the high-density
region. However, more information is required than available
at present, and quantitative predictions are limited.
%
%
\section{Summary}
In this paper we studied the deconfinement phase transition from hadronic
matter to a quark-gluon plasma as a function of the isospin.
The analysis was based on a separate description of the hadronic
and the quark-gluon phase.
We have demonstrated that in a consistent treatment of the two conserved
charges, baryon number and isospin, qualitatively new features arise.
Most importantly, the phase transition in asymmetric matter is generally 
continuous and of second-order. This is in contrast to the
discontinuous first-order behavior of symmetric systems, which arise
as singular points in the phase diagram of asymmetric matter.

The analysis was based on the determination of the phase separation
boundary, the binodal, which indicates the
region where a separation into two phases is energetically favorable.
The binodal is determined by Gibbs' criteria for phase
equilibrium. In a system with two conserved charges, it forms a two-dimensional
surface in parameter space, in contrast to the one-dimensional surface
in one-component systems.
The greater dimensionality leads to the qualitatively new behavior.
Primarily, the transition in asymmetric matter is continuous and of second
order. We saw that the pressure, temperature and isospin ratios in the
participating phases change throughout the transition. That means that in
general, asymmetric systems pass through a transition region rather than
staying at a transition point which leads to discontinuities in one-component
and symmetric systems.

To apply these results to the deconfinement phase transition in nuclear matter,
we employed a simple two-phase model.
The hadronic EOS was generated in a relativistic mean-field
model involving the interaction of baryons with isoscalar scalar and vector
fields and with the isovector $\rho$ meson and pion field.
This model involves cubic and quartic isoscalar self-couplings and
allows for an accurate calibration at normal nuclear densities to reproduce 
bulk properties of finite nuclei and nuclear matter.
For the description of the quark-gluon phase, we adopted a bag-model type
EOS involving massless $u$ and $d$ quarks.
In this framework thermodynamic quantities consists of two parts. A
perturbative contribution, where we included the first order correction 
in the strong coupling constant, and a nonperturbative contribution, represented
by the bag constant, which accounts for the QCD trace anomaly.
Guided by recent results from lattice
calculations, we chose the bag constant and the strong coupling constant to
reproduce a transition temperature of 150 MeV for matter at zero baryon
density.
However, the calibration of the hadronic EOS at normal densities is not
unique and predictions at high densities are sensitive to the model
parameters.
By studying variations of the quartic vector meson coupling within the natural
range, we found significant uncertainties in the predictions for the onset of
the deconfinement phase transitions in systems at finite baryon density.
For sufficiently large values of this coupling the hadronic EOS becomes so
soft that the transition vanishes altogether.

This two-phase model was then used to study the deconfinement phase transition
in highly excited matter that is 
produced in ultra-relativistic heavy-ion collisions.
By construction, an approach based on a separate description of the two phases
cannot reveal the actual nature of the transition.
However, until more concrete and reliable lattice calculations are available,
we believe that despite this limitation our analysis is useful for providing
a first orientation and concrete description of the qualitative features which
arise in the phase diagram of strongly interacting {\em asymmetric} matter.
There are several significant differences between the phase diagram
for an asymmetric system and that for symmetric matter.
In the first place, the location of the transition region in parameter space
depends on the isospin ratio of the system. Generally, the onset of phase
separation occurs at lower baryon densities and temperatures in more
asymmetric systems.
More importantly, the dimensionality of the phase-separation region
is larger in asymmetric matter leading to a {\em continuous} transition.
This implies that the thermodynamic properties of the participating
phases change throughout the transition.
Although these effects are small for realistic isospin ratios
$x\approx -0.1$, the trend in the isospin dependence to lower densities
and temperatures could be an indication that the transition region
is easier to reach in collisions of very neutron rich nuclei, which might
be created in radioactive beam facilities.

The analysis also revealed interesting physics in the isovector
channel. Most importantly, the incorporation of the pions into the model for
the hadronic phase was crucial for the shape of the binodal
surface and therefore also for the path of the system through the mixed-phase
region. The salient feature is the onset of a pion condensate at large baryon
densities and in very asymmetric systems.
This phenomenon is certainly of minor practical importance since the relevant
isospin ratios lie beyond the experimentally accessible region.
However, interest in Bose condensation has recently been revived in the
literature, particularly since the possibility of kaon condensation in neutron
star matter has been proposed \cite{POLITZER91}.
Analyzing these effects in connection with the deconfinement phase transition
provides an important topic for future work on this problem.

\acknowledgments

I am pleased to thank B. D. Serot for many useful comments and stimulating
discussions.
This work was supported by the U.S. Department of Energy under
contract No.~DE-FG02-87ER40365.
\newpage
%
%
%
\begin{table}[tbhp]
\caption{Equilibrium Properties of Nuclear Matter}
\medskip
\begin{tabular}[b]{cccccc}
$k_{\scriptscriptstyle\rm F}^0$ &
		 $\rho_B^0$ 
                & $M_0^* /M$ &
                $e_{\scriptscriptstyle 0}$ & $K_0$ & $a_4$ \\
\hline
1.30\,fm$^{-1}$ & 0.1484\,fm$^{-3}$ & 0.60 & $-15.75$\,MeV & 250\,MeV &
					 35\,MeV\\
\end{tabular}
\label{tab:one}
\end{table}
\newpage
%
%

%
\newpage
%
%
\section*{Figure captions}
\global\firstfigfalse
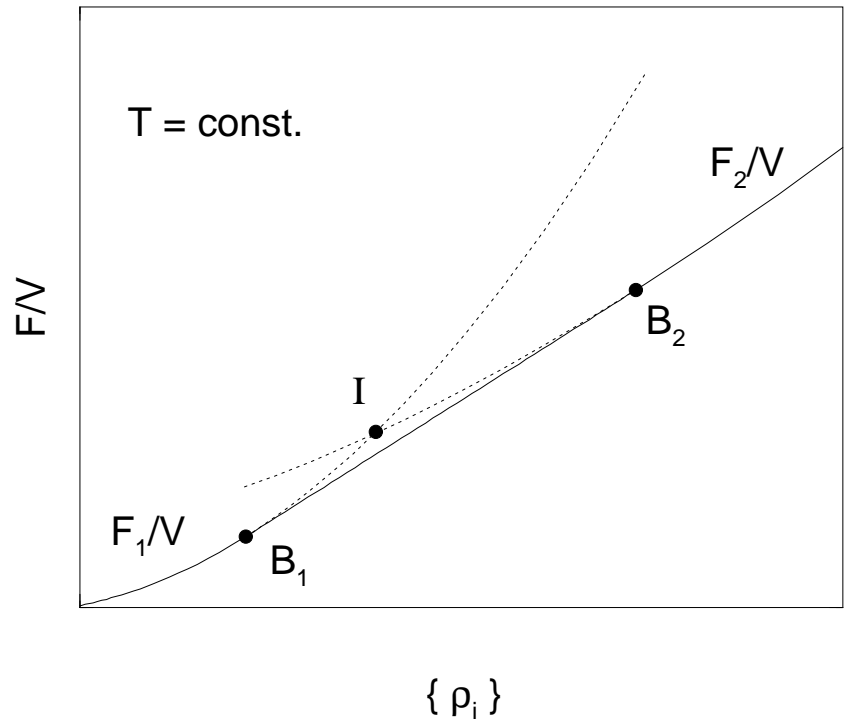
\begin{figure}
\caption{
The Maxwell construction for a system with more than one conserved charge.
The free energy density as a function of the densities $\{\rho_i\}$ at
constant temperature is shown.
The Maxwell construction is indicated by the segment $\overline{B_1B_2}$.
The course of the free energy in each single phase is denoted by the dotted
lines.
The metastable configurations are located on $\overline{B_1I}$ and 
$\overline{B_2I}$ between the binodal and the point where the free energy
curves intersect.
}
\label{fig:max}
\end{figure}
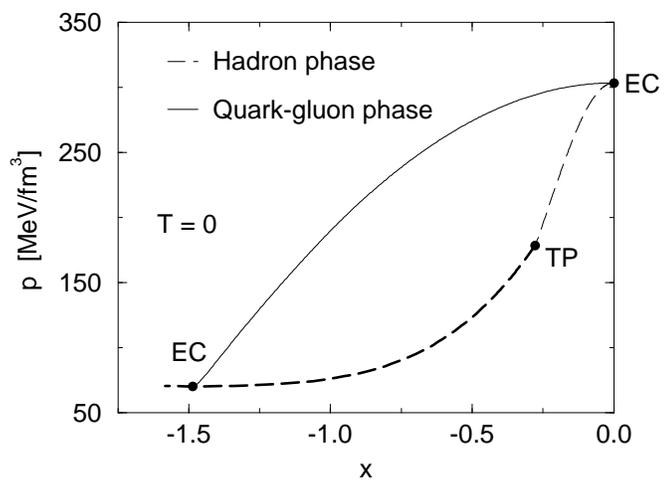
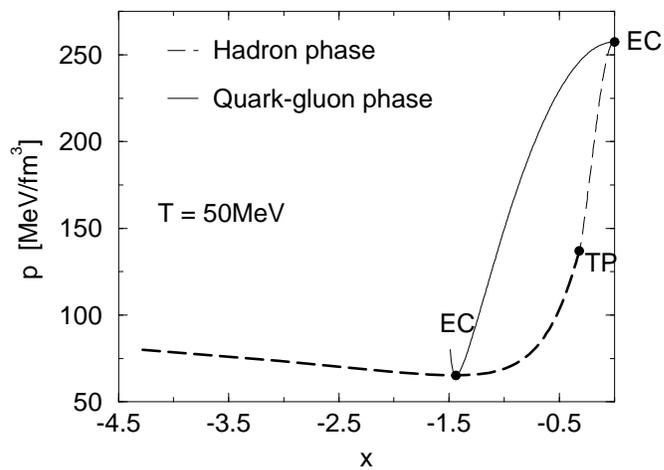
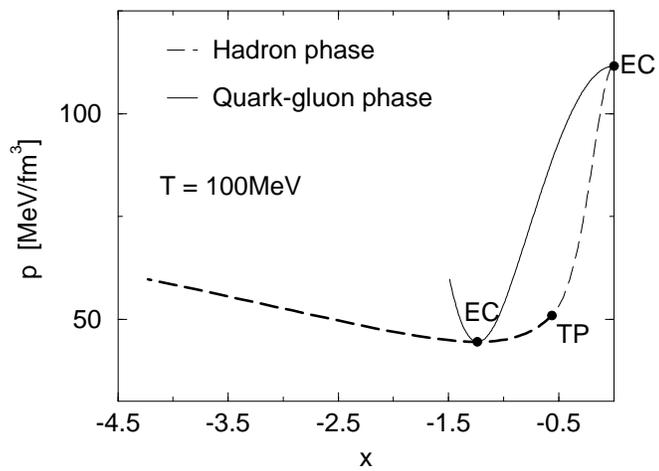
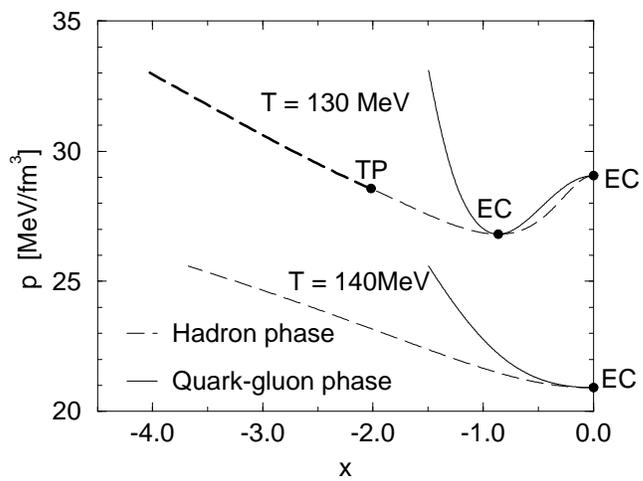
\begin{figure}
\caption{
The binodal surface indicating the two-dimensional phase-coexistence
boundary in $(p, T, x)$ space.
A sequence of binodal sections at fixed temperature is shown.
The points of equal concentrations (EC) and the points of three phases (TP)
are indicated.}
\label{fig:slices}
\end{figure}
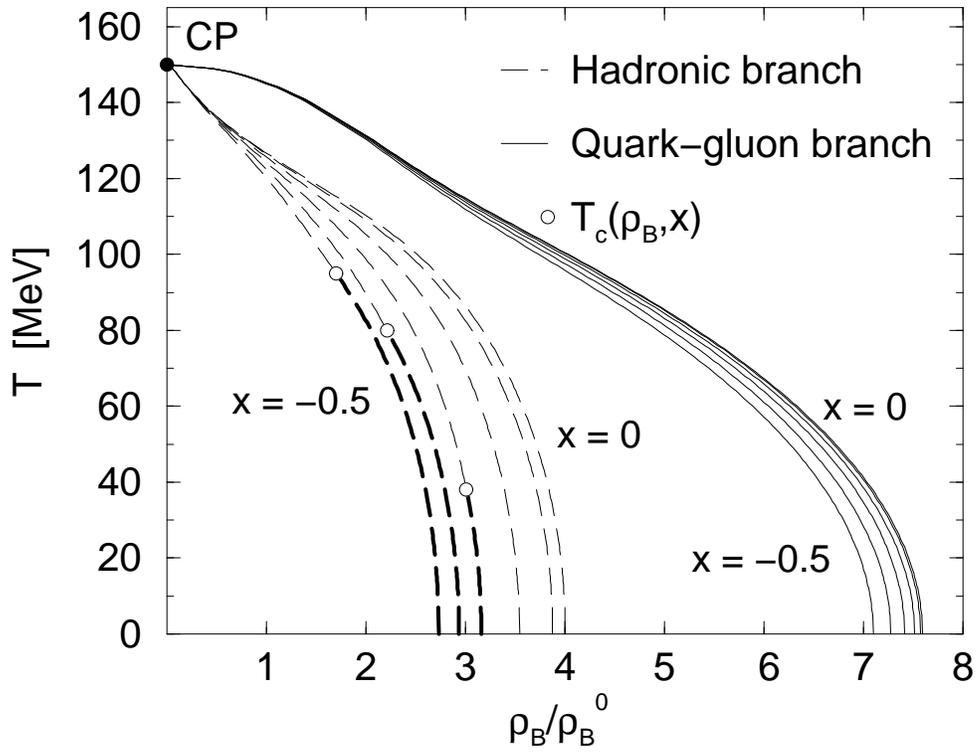
\begin{figure}
\caption{
Binodal section in $(\rho_B, T, x)$ space.
The projection of several binodal branches at different $x$ onto the 
$(\rho_B,T)$ plane is shown. The isospin ratios are
$x= 0, -0.1, -0.2, -0.3, -0.4, -0.5$, starting from the right.
The open circles mark the critical temperature for the onset of the pion
condensate.
}
\label{fig:bintvsrho}
\end{figure}
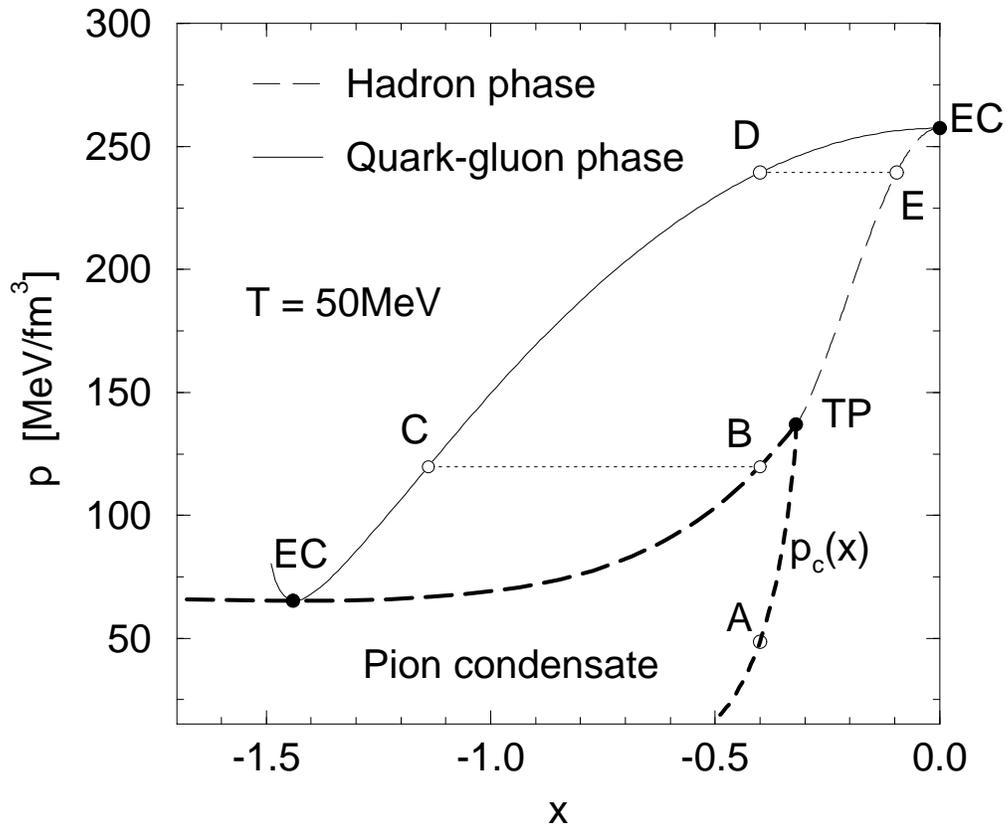
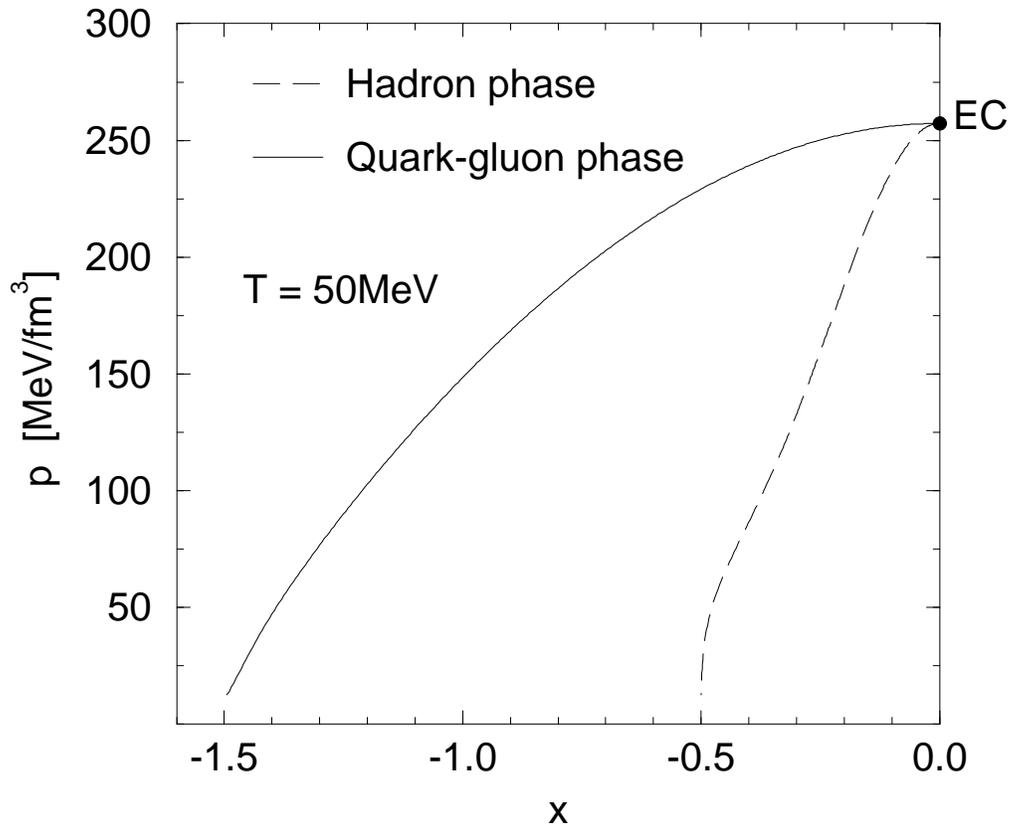
\begin{figure}
\caption{
(a) Binodal section at $T=50\,{\rm MeV}$ including the contributions of 
the pions in the hadronic phase.
The points $B$ through $E$ denote phases participating in a phase
transition. At $A$ the system crosses the critical pressure curve which
marks the onset of pion condensation.
The point of three phases (TP) and the points of equal concentration (EC)
are also indicated.
(b) Binodal section at $T=50\,{\rm MeV}$ without the pionic contributions.
}
\label{fig:slice}
\end{figure}
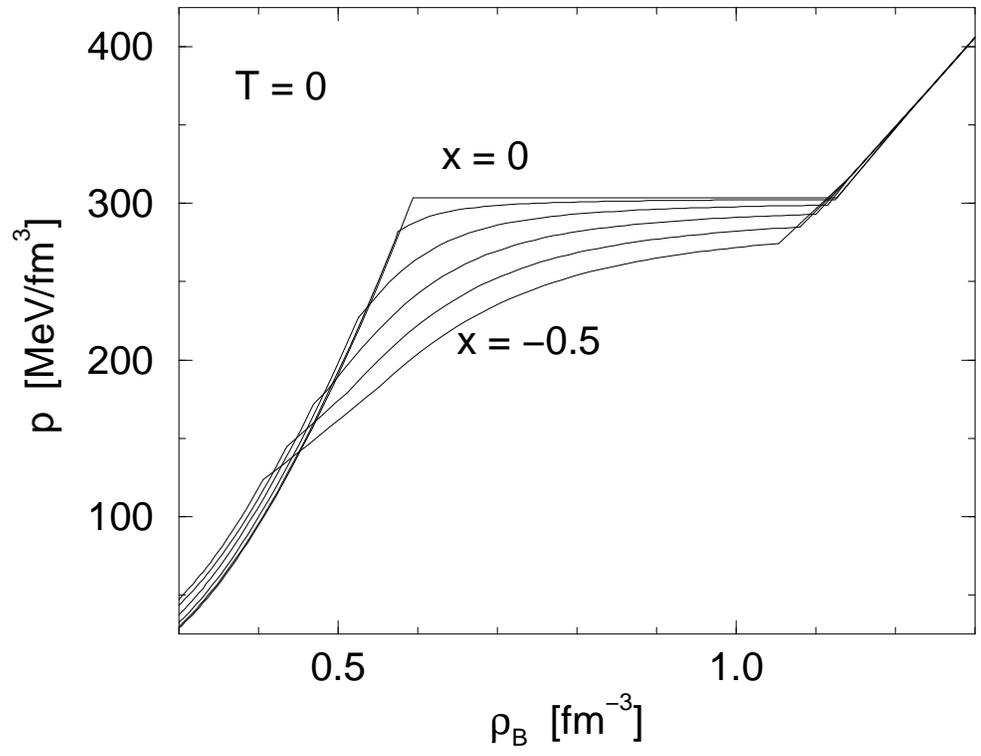
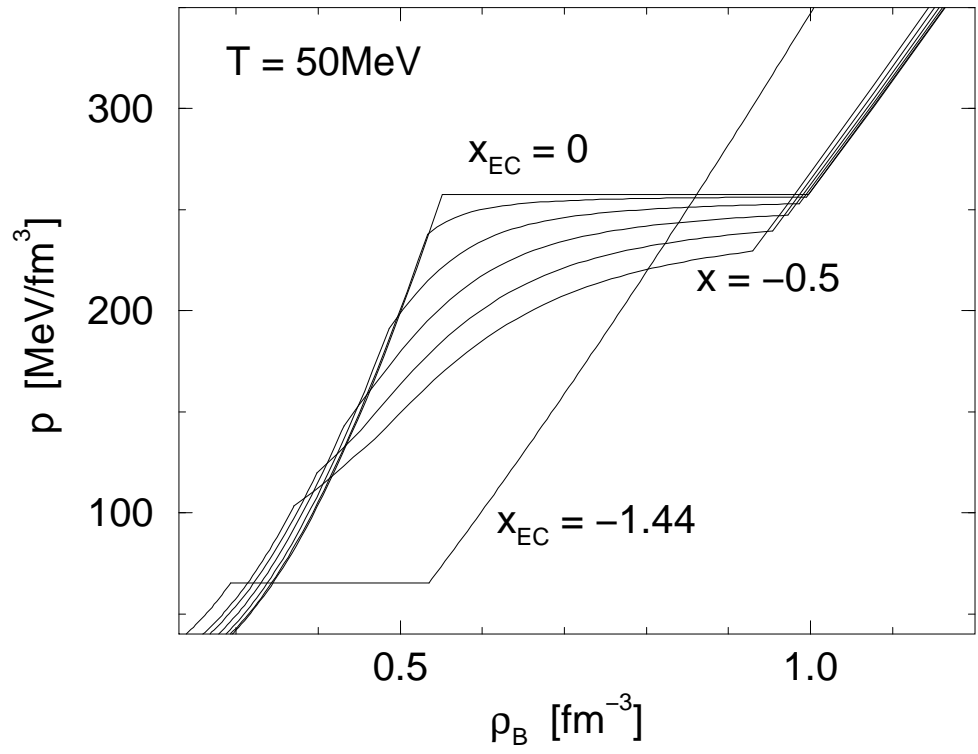
\begin{figure}
\caption{(a) Isotherms for different values of $x$ at $T=0$.
The isospin ratios are $x=0, -0.1, -0.2, -0.3, -0.4, -0.5$ from the top to the
bottom.
(b) Isotherms for different values of $x$ at $T=50\,{\rm MeV}$.
The isospin ratios are the same as in part (a). In addition the isotherm with
$x_{\scriptscriptstyle EC}=-1.44$, which corresponds to the second point of
equal concentration, is also indicated.
}
\label{fig:pvsrho}
\end{figure}
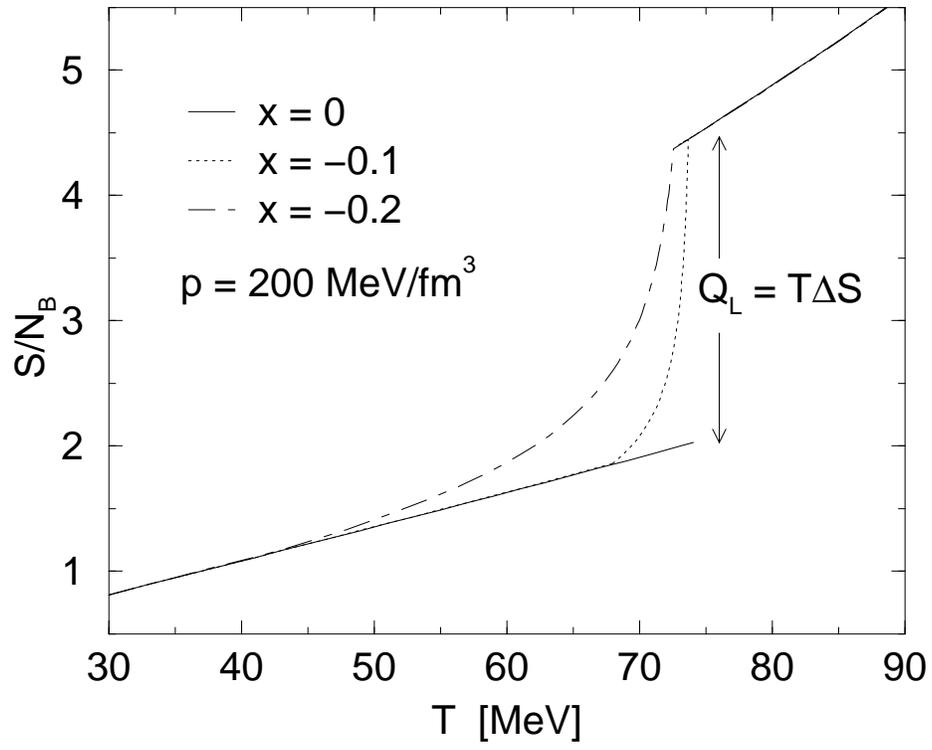
\begin{figure}
\caption{
Entropy per baryon as a function of temperature at constant pressure
for various isospin ratios.
}
\label{fig:svst}
\end{figure}
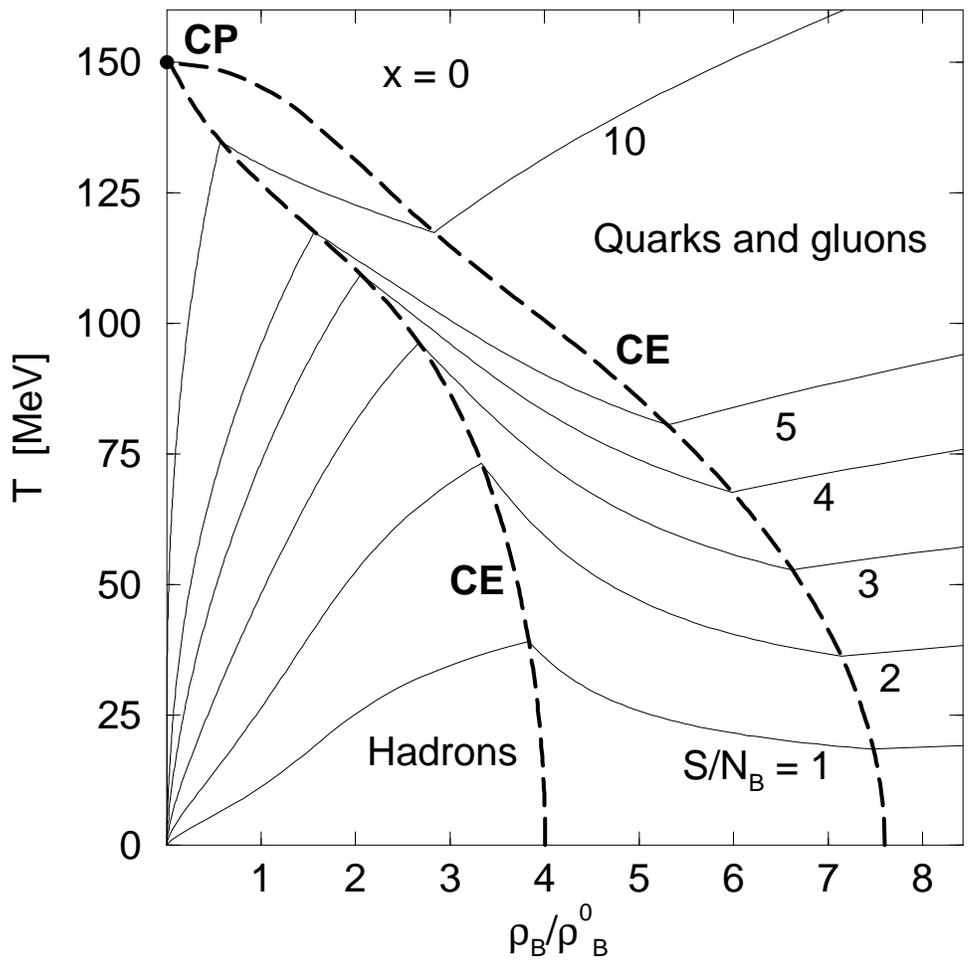
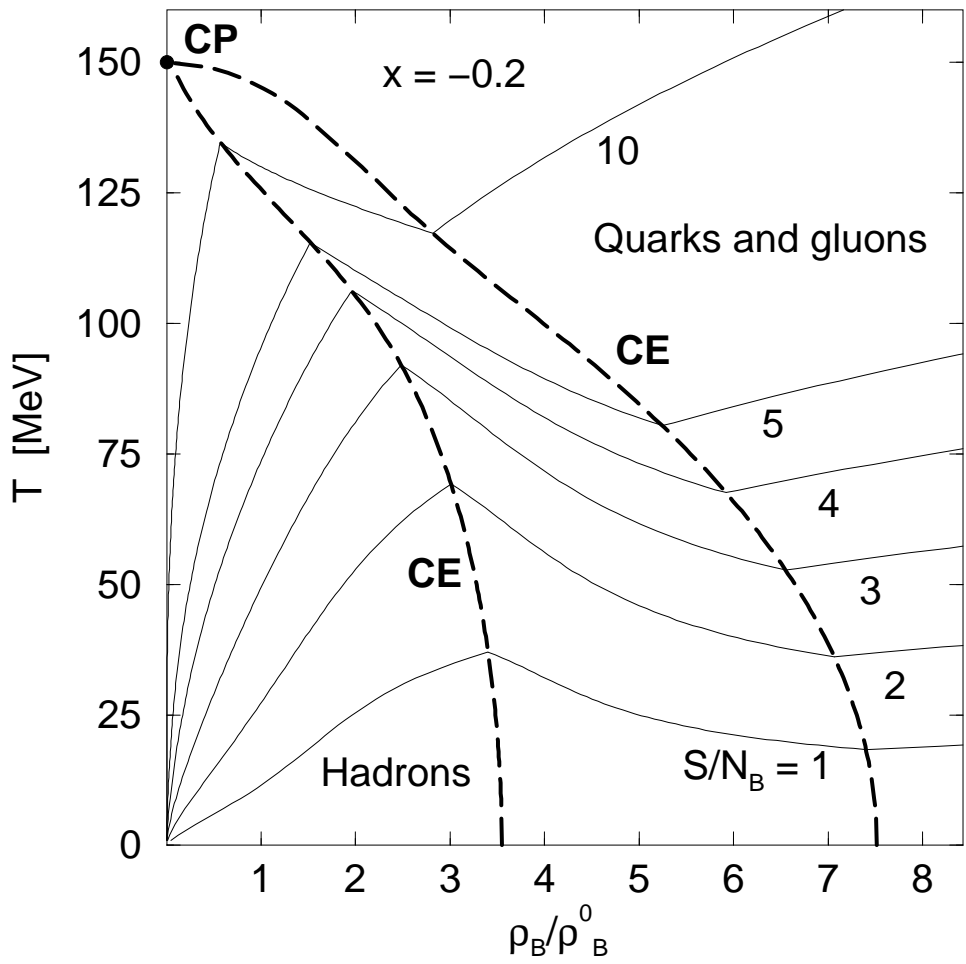
\begin{figure}
\caption{
Properties of nuclear matter as functions of temperature and density.
The dashed lines denote the coexistence curves (CE), which intersect at the
critical point (CP). The solid lines indicate adiabats.
The lines between the coexistence curves are the stable mixed-phase
configurations.
Part (a) shows results for symmetric matter ($x = 0$) and part (b), 
for asymmetric matter  with $x=-0.2$.}
\label{fig:tvsrhos}
\end{figure}
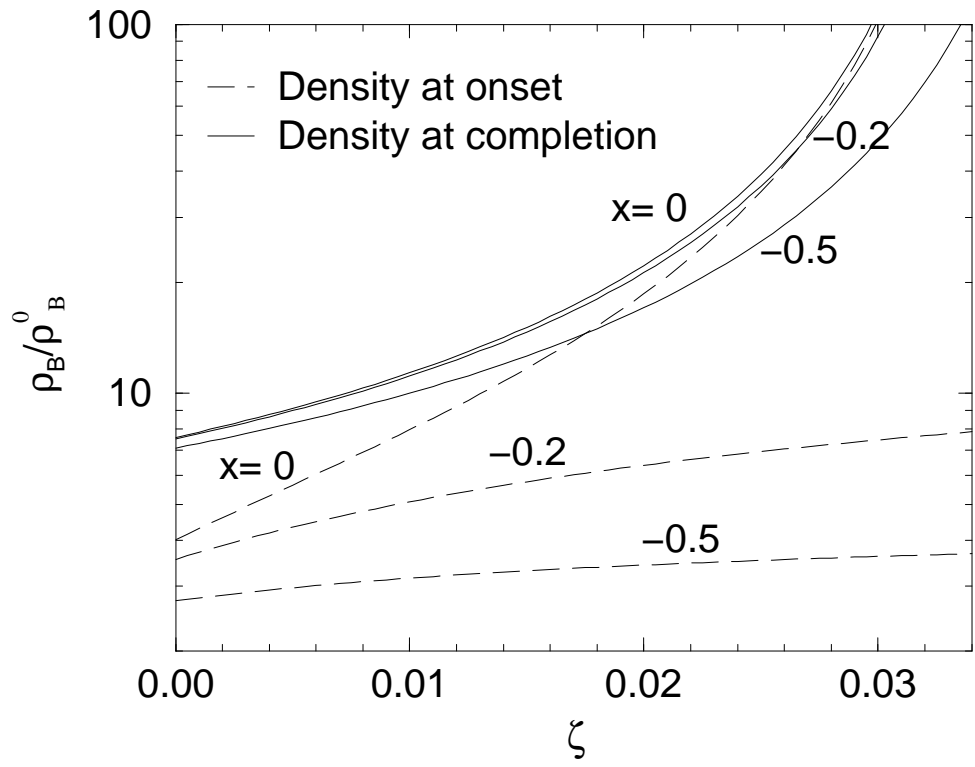
\begin{figure}
\caption{Baryon density at the onset and at the end of the
transition as a function of the quartic vector meson coupling $\zeta$.
}
\label{fig:rhovszeta}
\end{figure}
\end{document}